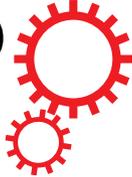

# SCIENTIFIC REPORTS

# A neutron tomography study: probing the spontaneous crystallization of randomly packed granular assemblies

Indu Dhiman[1], Simon A. J. Kimber[4], Anita Mehta[3] & Tapan Chatterji[2]

We study the spontaneous crystallization of an assembly of highly monodisperse steel spheres under shaking, as it evolves from localized icosahedral ordering towards a packing reaching crystalline ordering. Towards this end, real space neutron tomography measurements on the granular assembly are carried out, as it is systematically subjected to a variation of frequency and amplitude. As expected, we see a presence of localized icosahedral ordering in the disordered initial state (packing fraction ≈ 0.62). As the frequency is increased for both the shaking amplitudes (0.2 and 0.6 mm) studied here, there is a rise in packing fraction, accompanied by an evolution to crystallinity. The extent of crystallinity is found to depend on both the amplitude and frequency of shaking. We find that the icosahedral ordering remains localized and its extent does not grow significantly, while the crystalline ordering grows rapidly as an ordering transition point is approached. In the ordered state, crystalline clusters of both face centered cubic (FCC) and hexagonal close packed (HCP) types are identified, the latter of which grows from stacking faults. Our study shows that an earlier domination of FCC gives way to HCP ordering at higher shaking frequencies, suggesting that despite their coexistence, there is a subtle dynamical competition at play. This competition depends on both shaking amplitude and frequency, as our results as well as those of earlier theoretical simulations demonstrate. It is likely that this involves the very small free energy difference between the two structures.

Granular packings of dry hard spheres have been widely studied, in keeping with their wide-ranging applications[1–9]. These[3,10,11] have been informed to some extent by the modelling of various disordered systems like glasses[1,5,12], liquids[13] and colloidal systems[14–18], in terms of structural evolution, jamming states, dynamical arrest and mechanical rigidity. The origins of the field were pioneered by Bernal's experiments in the 1960s[19,20], where monodisperse spherical particles were packed compactly by shaking the container in which they were placed. The resulting packings of hard spheres were characterized by the presence of densely packed polytetrahedral clusters. Although these polytetrahedral aggregates favor relatively dense packing, their incompatibility with translational symmetry precludes periodic structures leading to the emergence of geometric frustration. These findings echoed the hypothesis presented in 1952 by Frank et al.[21] that the presence of local icosahedral clusters might inhibit crystallization, the icosahedral clusters allowing relatively dense packings without translational symmetry. A similar finding on the role of icosahedral clusters in stabilizing the disordered state was reported by Steinhardt et al.[22] using computer simulation methods. In 1989, Edwards et al.[23] provided a theoretical framework for these ideas and explored analogous behavior based on thermodynamics to understand granular packings. Thereafter, several studies have been reported towards the understanding of five-fold symmetric structures in liquids and amorphous materials[1–19]. It is important to mention that 20 tetrahedral subunits can form an icosahedral structural configuration, therefore the terms 'polytetrahedral' and 'icosahedral' have been used interchangeably in the literature. These tetrahedral subunits are also known to form other types of structural configurations.

[1]Neutron Scattering Division, Oak Ridge National Laboratory, Oak Ridge, Tennessee, 37831, USA. [2]Institut Laue-Langevin, 71 Avenue des Martyrs, 38000, Grenoble, France. [3]Max Planck Institute for Mathematics in the Sciences, Inselstrasse 22, 04103, Leipzig, Germany. [4]Université Bourgogne-Franche Comté, Université de Bourgogne, ICB-Laboratoire Interdisciplinaire Carnot de Bourgogne, Bâtiment Sciences Mirande, 9 Avenue Alain Savary, B-P. 47870, 21078, Dijon Cedex, France. Correspondence and requests for materials should be addressed to I.D. (email: dhimani@ornl.gov) or T.C. (email: chatterji@ill.fr)

Relatively recent simulation studies reported by Anikeenko et al.[24,25] emphasize that polytetrahedral structure is a key feature of disordered packing. They suggest that the fraction of polytetrahedral aggregates grows with increase in disordered packing fraction up to the random packing fraction limit, $\phi = 0.64$. This study, in addition to other studies reported in the literature[20,26,27], also underlines the preference of polytetrahedral over icosahedral configurations. Francois et al.[28] and co-workers[29–31] investigated the crystallization of granular sphere packings experimentally. They observed that the densification of the disordered phase is correlated with an increase in the fraction of polytetrahedral clusters[27], the fraction of which again decreases with the development of crystallinity, in agreement with Anikeenko et al.[24,25] Mehta et al.[3,10], and[11] have also carried out extensive work in the field of granular media using three-dimensional Monte Carlo simulations, in which a spontaneous transition from disordered to crystalline ordered states was demonstrated for the first time[32]. Further, Shinde et al.[33] suggested the existence of an optimal range of amplitudes at which complete crystallization could occur, while also observing competition between face centered cubic (FCC) and hexagonal close packed (HCP) ordering.

The structural transition from disordered to partially crystalline packings is known to involve a wide variety of mechanical, geometrical and topological structures. In particular, the disordered packing fraction is known to be limited to a maximum of $\phi = 0.64$, i. e. the packing fraction can only increase beyond this value upon the appearance of crystalline clusters. Despite significant progress in the field, a complete understanding of the structural evolution of dry hard sphere packings and their transition from a disordered state towards a crystalline state is still lacking. The characterization of the many competing interactions present in such dense packings with the frequent appearance of geometrically frustrated structures could thus be central to the understanding of the spontaneous evolution of crystallinity emerging from disorder. Given that granular systems are athermal[3], mechanical perturbations[8,9,34] in the form of shaking, vibrating, shearing or tapping are needed to observe such and similar relaxation behavior.

Pioneering experimental studies on the compaction behavior of granular materials as a function of mechanical shaking were carried out by Knight et al.[9] and Nowak et al.[35]. In these studies, loosely packed beads were confined in a thin tube, which was then tapped vertically. As a function of vertical tapping, very slow compaction of the system was observed, whose relaxation behavior could be described based on an inverse – logarithmic fit. However, these experiments were limited by wall effects, which favored the ordering of the granular material along the container walls. Philippe et al.[34,36] and Ribière et al.[37,38] also studied the compaction dynamics of granular materials as a function of tapping, taking into account that ordering due to wall effects was negligible.

With the substantial advancement of 3D tomography techniques (both with x-rays and neutrons), it has become possible to explore structures and their evolution in three-dimensional granular packings. Yet, not many detailed experimental studies have been reported in this direction. The first x-ray tomography based experiment on granular matter was reported by Seidler et al.[39]. This study highlighted the importance of tomography measurements to investigate the 3 – dimensional structural evolutions in nominally monodisperse glass spheres. These studies were further extended by Richard et al.[40], where granular particles were vertically vibrated to study the packing structure, mainly the volume distribution, and gradual densification of the system. Detailed x-ray tomography studies with elaborate local structural analysis (such as calculation of pair correlation functions, bond orientational order, contact network and local free volume) have been presented by Aste et al.[41–43] According to their studies, the evolution of quasi-perfect tetrahedral structures is accompanied by a decrease in entropy. Francois et al.[28] also utilized the x-ray microtomography experimental technique to investigate the emergence of geometrically frustrated structures in granular sphere packing systems, as discussed above. Recently, Xia et al.[44,45] performed systematic studies using x-ray tomography to investigate structural heterogeneities and their evolution as well as the dynamical behavior and related thermodynamics of granular packings, and suggested that geometrically frustrated polytetrahedral order could be utilized as a 'glassy' structural order parameter. Furthermore, a non-cubic scaling law[45] was observed for disordered granular spheres similar to metallic glasses, which the authors correlated with a jamming transition[46].

In addition to X-ray tomography methods, there are also other experimental techniques used to study the problem of granular packing, such as confocal microscopic imaging, refractive index matched imaging and magnetic resonance imaging (MRI). A detailed comparison of these techniques has been reviewed by Amon et al.[47] and Dijksman et al.[48].

Neutron tomography is uniquely suitable for the investigation of an assembly of steel spheres, owing to the high penetration depth inherent to neutrons. This also allows for the use of large sample containers, to better avoid wall effects. Also, we note that steel spheres are much more monodisperse than the glass spheres typically used for x-ray investigations. In the present work, we have carried out systematic neutron tomography measurements on hard steel spheres, as a function of the frequency and amplitude of shaking. The fact that the cylindrical container is made of aluminum ensures that the spheres are not electrically charged; the non – magnetic nature of the spheres at room temperature substantially rules out inter-particle magnetic forces. The only interaction between the spheres is hard-core repulsion, which is invoked when spheres touch each other. Using neutron tomography, the position of each individual hard sphere can be determined. By varying the frequency and amplitude of shaking we can thus explore the morphological evolution of the (partially) crystalline state from the randomly packed state.

Neutron tomography enables us to probe structure at short as well as medium ranges, enabling us to explore the effects of both geometrical frustration as well as the onset of crystallinity. Our results suggest that while icosahedral ordering is localized, the crystalline ordered phase grows upon approaching the transition point.

## Experimental Details

Neutron tomography measurements were carried out at the CG1-D imaging beam line using cold neutrons, located at the High Flux Isotope Reactor (HFIR) at Oak Ridge National Laboratory[49]. The schematic of the experimental set-up is shown in Fig. 1.

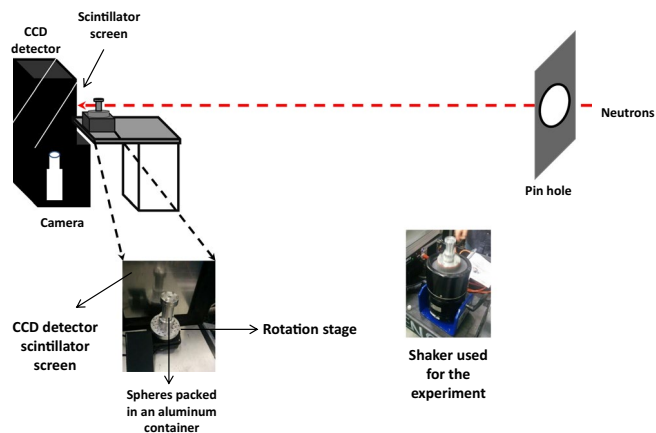

**Figure 1.** The schematic of the experimental set-up for the neutron imaging measurements at CG-1D instrument at ORNL HFIR.

Neutron attenuation contrast from steel spheres was observed using a charge coupled detector (CCD) camera system (iKon – L 936, Andor Technology plc. Belfast, UK), with 25 μm thick LiF/ZnS scintillator, the field of view (FOV) ≈ 7.4 × 7.4 cm, and the effective pixel area of 36 × 36 μm². The cold neutrons had a wavelength range from 0.8 to 6 Å, with a peak intensity of $2.2 \times 10^6$ n/cm²/s at 2.6 Å. The aperture to detector distance was L = 6.59 m, with an aperture size D = 16 mm. The obtained resolution was ≈100 μm.

To carry out these measurements, steel spheres of 1 mm (±0.001 mm) diameter were poured into a cylindrical Al container with an inner diameter of 25 mm. To measure the initial state (randomly closed packed, with packing fraction $\phi \approx 0.62$) these steel spheres were poured slowly in the container. Thereafter, to study the structural evolution from a randomly closed packed to a crystalline state the sample container was vertically shaken for five minutes, at various frequencies for 0.2 and 0.6 mm amplitude values. For 0.2 and 0.6 mm shaking amplitudes, tomography measurements were performed at 40, 85, and 180 Hz. For every combination of frequency and amplitude, the sphere configurations were reinitialized by slow pouring into the container. To perform neutron tomography measurements, the container was gently mounted on a rotation stage after each shaking cycle. The exposure time for each radiograph/projection was 50 secs and the measurement was approx. 10–12 hours. The 3D reconstruction for each data set, mapping the linear absorption coefficient in the steel spheres, was carried out with commercially available OCTOPUS software package using the standard filtered back-projection algorithm[50].

Subsequently, the location of each particle was acquired by segmenting the 3D data set using AMIRA software. To minimize wall effects, we have focused only on the central part of the cylinder and have removed the data corresponding to about four particle diameters from the container walls in the analysis. The OVITO[51] software was used to obtain the radial distribution function, angular distribution function, and to perform the Voronoi tessellation and adaptive – common neighbor analysis (ACNA)[52].

## Results and Discussion

Figure 2 shows representative 3D projections of the segmented spheres in XY, YZ – planes and in 3D, with Fig. (2a–c) reflecting the randomly packed initial state corresponding to spheres gently poured into the container. Figure (2d–f) reflect the relatively ordered configurations corresponding to shaking at a frequency of 180 Hz, and an amplitude of 0.6 mm. In Fig. (2a–c) an overall randomly packed structure (with packing fraction $\phi \approx 0.62$) can be clearly observed. This value of fraction is close to the random-close-packed fraction $\phi \approx 0.64$ at which spheres can pack densely without crystalline ordering[19]. After shaking at a frequency of 180 Hz and an amplitude of 0.6 mm, a transition to a crystalline state occurs with a concomitant increase in packing fraction to $\phi \approx 0.71$, as observed in Fig. (2d–f). The variation of packing fraction as a function of frequency, for 0.2 and 0.6 mm shaking amplitudes is shown in Fig. 3. Note that while there is also an increase of packing fraction (to a maximal value of $\phi \approx 0.66$) in the graph corresponding to the 0.2 mm shaking amplitude, this is far less pronounced than what is exhibited at the higher shaking amplitude of 0.6 mm. A first-order-like transition to a packing fraction of $\phi \approx 0.71$ is reached (to be compared with the densest possible packing fraction of $\phi \approx 0.74$). This is in conformity with the predictions of Shinde et al.[33]. who showed that spontaneous crystallization to FCC and HCP occurs more completely at higher amplitudes within a specified range.

Next, we study the spatial correlations as a function of frequency; our results are presented in Fig. 4(a,b), where the pair distribution functions (PDF) $G(r)$ of steel spheres are plotted as a function of frequency for amplitudes of 0.2 and 0.6 mm, respectively. The first peak at $r \approx 1$ (diameter of the spheres) is the most pronounced, corresponding to nearest neighbors in contact. Note the sharp leading edge in all of the plots, which implies that the vast majority of spheres are in contact with their nearest neighbors. Thereafter, the probability of finding the neighbors is reduced. For frequency values between 0 and 85 Hz, the $G(r)$ in Fig. 4(a) displays two sub-peaks at $r = \sqrt{3}$ and 2, the next near – neighbor regions. This split in the second peak into two sub – peaks is an attribute of randomly packed spheres, signifying the strong local order in the first two coordination shells. The first sub – peak at $r = \sqrt{3}$ is due to the edge shared in–plane triangle/hexagonal packing, while the second sub – peak at $r = 2$ corresponds to a linearly aligned series of three or more spheres. As the frequency and the correlated packing fraction increases, both peaks increase in height, with the peak at $r = \sqrt{3}$ showing faster growth, suggesting that the number of contacts increase. This behavior may indicate that number of configuration in the contact network increases. On further increase in frequency to 180 Hz, another peak at r = $\sqrt{2}$ emerges. The emergence of such distinct sub – peaks and more peaks at higher r values with an increase in packing fraction signifies the development of crystalline ordering, even though no clear signature of complete crystallization is observed. This behavior is discussed in greater details based on Voronoi tessellations and the ACNA analysis described below. In Fig. 4(b), we plot G(r) as a function of r for a shaking amplitude of 0.6 mm and several values of frequency. At 40 Hz, G(r) displays signatures similar to that seen in Fig. 4(a) for 85 Hz at 0.2 mm amplitude, suggesting the earlier appearance of local ordering. With increasing frequency beyond this point, more sharp peaks appear, indicating a more rapid evolution towards crystallinity. In particular, the sub-peaks at $r = \sqrt{2}, \sqrt{3}$ and 2 appear to be much more pronounced, as compared to the ones observed for a shaking amplitude of 0.2 mm (Fig. 4(a)). This suggests a higher degree of crystalline ordering.

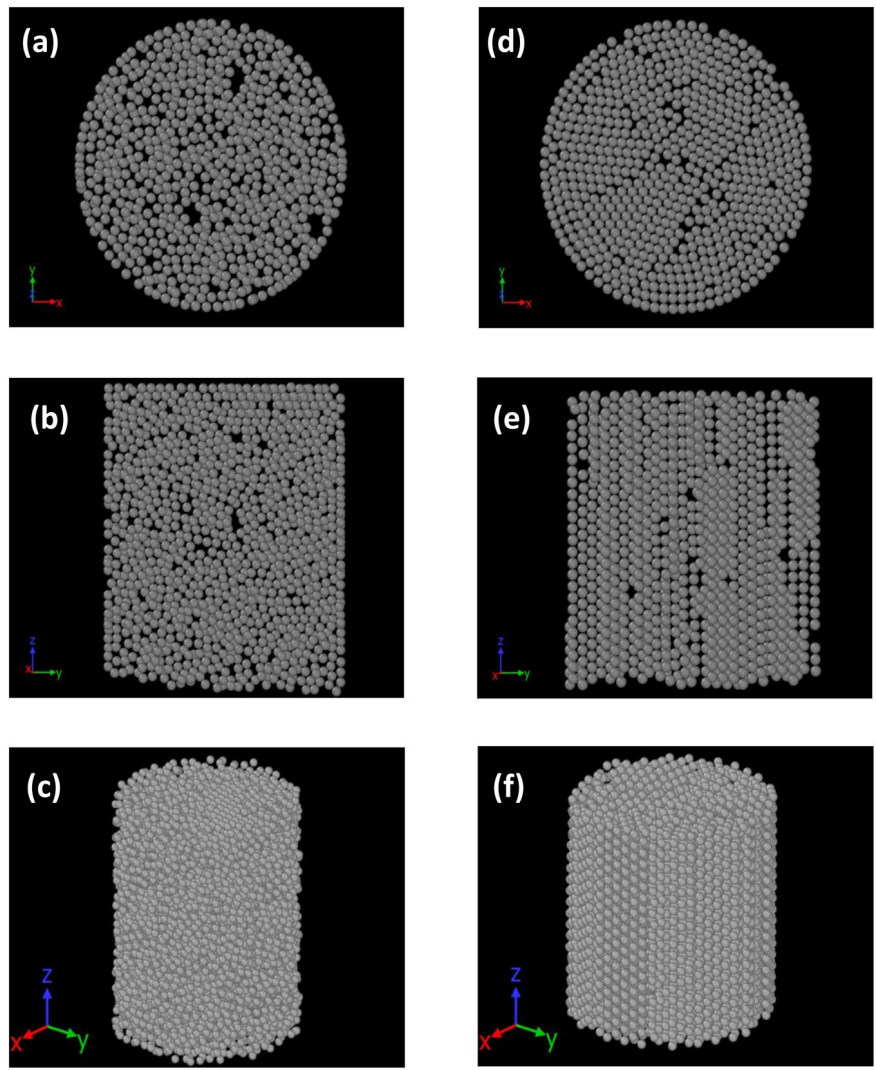

**Figure 2.** Representative 3D projections of the segmented spheres (diameter = 1 mm) inside a cylindrical container (inner diameter = 25 mm) in the initial state in (**a**) XY, (**b**) YZ – planes (**c**) 3D and at 180 Hz, 0.6 mm shaking amplitude in (**d**) XY, (**e**) YZ – planes and (**f**) 3D.

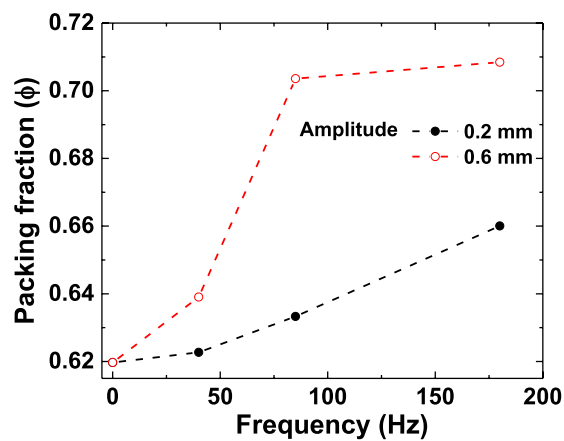

**Figure 3.** The variation of packing fraction as a function of frequency for 0.2 and 0.6 mm shaking amplitudes.

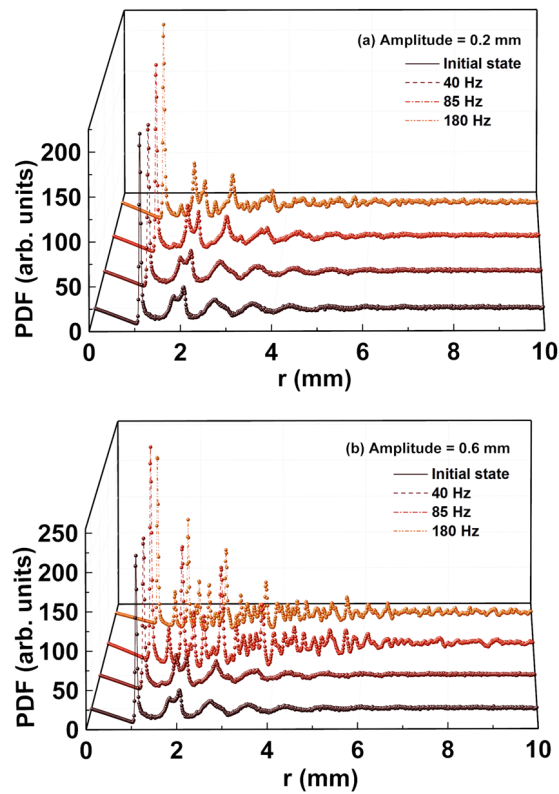

**Figure 4.** Pair distribution function (PDF), G(r), plots, at various frequency values for shaking amplitudes of (**a**) 0.2 mm and (**b**) 0.6 mm.

The pair distribution function G(r) discussed above thus clearly shows the emergence of higher order structural correlations with a transition from random to crystalline local ordering, as a function of frequency and amplitude. In order to move beyond such two-body correlations to look at more global features, we need to calculate higher-order structural correlations from the data.

Information about three-body structural correlations can be deduced from the angular distribution function presented in Fig. 5(a,b), where angular distribution functions (ADF) versus angle (θ) for 0.2 and 0.6 mm amplitudes are respectively plotted. The ADF is defined as the probability distribution of the angle (θ) between two spheres, which are touching the same sphere. In the initial state, the ADF displays a sharp peak at θ ≈ 60° and a relatively broad peak at about ≈117°. The location of these two peaks is significantly close to the ideal values for the icosahedral structure observed at 63.5° and 116.5°, indicating the presence of distorted icosahedral ordering. The peak close to 60° indicates the presence of tetrahedra, which can form an icosahedral structural configuration. Interestingly, the broad peak at θ ≈ 117° exhibits an increase in height and a shift towards a higher angle of θ ≈ 120° at higher frequency values in Fig. 5(a). The frequency dependence of the angular shift and intensity variation of the peak from icosahedral (θ ≈ 116.5π) towards FCC/HCP (θ ≈ 120π) crystalline evolution for 0.2 mm shaking amplitude is shown in the inset to Fig. 6(a). This may indicate a structural evolution towards a diamond-like structure, as reported in[48]. In addition, at higher frequency values ranging from 85–180 Hz, another peak at θ = 90° emerges. This is a peak shared between the ideal FCC – type (peaks at θ = 60°, 90° and 120°), and the ideal HCP – type structure (peaks at θ = 60°, 90°, 108°, 120°, and 146°). Therefore, the emergence of a peak at θ ≈ 90° in Fig. 5(a) may indicate the growth of FCC/HCP – type structural correlations.

Similarly, in Fig. 5(b) the ADF at the higher amplitude of 0.6 mm also displays signatures of structural evolution from a disordered state towards a crystalline state. At the lower frequency value of 40 Hz, a sharp peak at θ ≈ 60° and a broad peak at θ ≈ 117° are observed, correlated with the presence of tetrahedral structural configurations. Similar to the case of the 0.2 mm shaking amplitude, the peak at θ ≈ 117° shifts towards higher θ ≈ 120π with increasing frequency values (shown in the inset to Fig. 6(b)), also for the 0.6 mm shaking amplitude. A further increase in frequency to 85 Hz favors the emergence of additional peaks at θ ≈ 90°, 109°, 120°, and 147°. The emergence of peaks at 109° and 147° is in contrast with the ADF for the shaking amplitude of 0.2 mm

  5

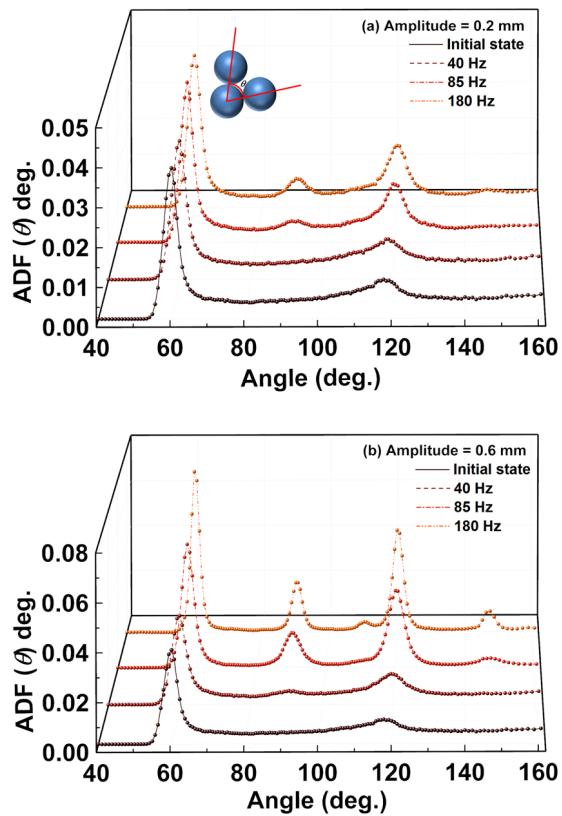

**Figure 5.** Angular distribution function, ADF(θ) plots, at various frequency values for (**a**) 0.2 mm and (**b**) 0.6 mm shaking amplitudes.

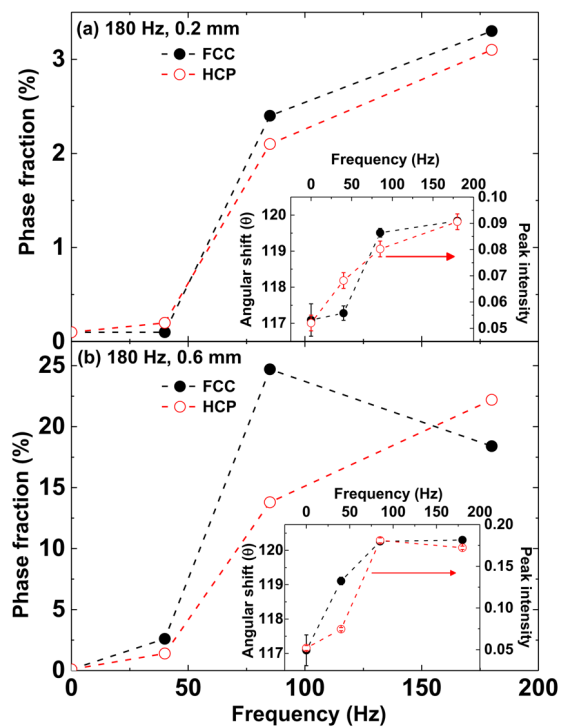

**Figure 6.** Phase fraction of face centered cubic (FCC) and hexagonal close packed (HCP) structures as a function of frequency for (**a**) 0.2 mm and (**b**) 0.6 mm shaking amplitudes. The insets to (**a,b**) show the frequency dependence of the angular shift and intensity variation of the peak from icosahedral (θ ≈ 116.5°) towards FCC/ HCP (θ ≈ 120°) crystalline evolution for 0.2 mm and 0.6 mm shaking amplitudes, respectively.

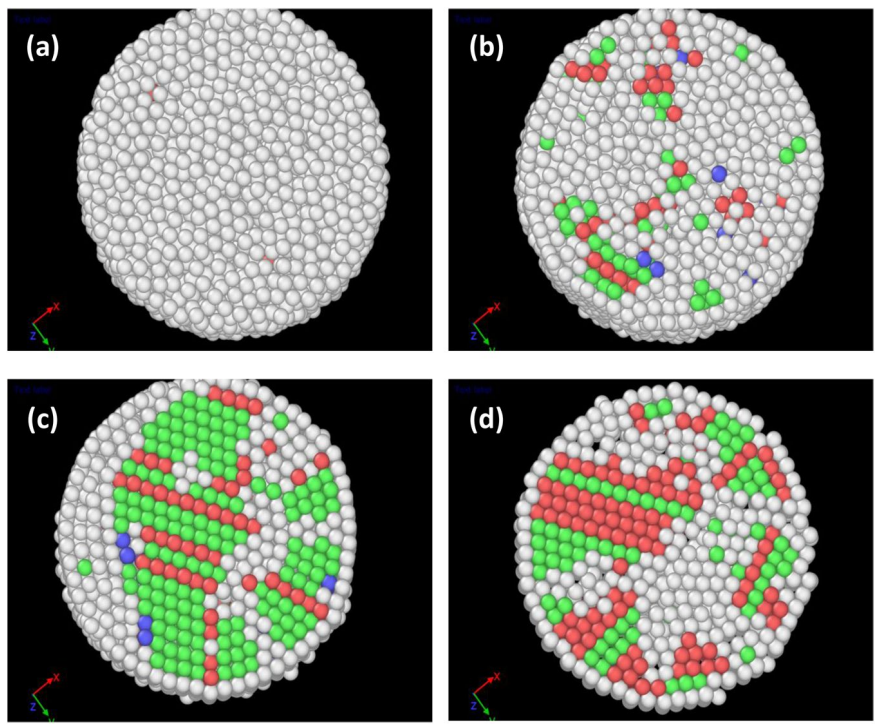

**Figure 7.** The structural evolution of different phases obtained using adaptive common neighbor analysis at 0.6 mm shaking amplitude for (**a**) initial, (**b**) 40 Hz, (**c**) 85 Hz and (**d**) 180 Hz frequency values. Green and red colored spheres represent face-centered-cubic and hexagonal-close-packed clusters, respectively. The blue ones correspond to body centered cubic type and the white colored spheres are the ones with undefined coordination, where the local environment of the spheres may not resemble any of the known crystal structures close enough. A video showing the structural evolution at 180 Hz frequency and 0.6 mm shaking amplitude value is given in the supplementary information.

(Fig. 5(a)). These two peaks have positions which are close to those observed for the HCP structure. These peaks become sharper and increase in height at the higher frequency of 180 Hz, which suggests that there has been an augmentation of HCP – type crystallinity.

Various approaches exist to characterize local structures, such as common neighbor analysis (CNA)[53], ACNA[51], bond angle analysis[54], Voronoi analysis[55], and polyhedral template matching methods[56]. In this work, the most commonly used methods, ACNA, bond angle analysis, and polyhedral template matching methods were employed and gratifyingly, all the results for the quantities we examined showed similar trends as a function of frequency. Here, we report only the results obtained from the ACNA method. The ACNA is an extension of the frequently used CNA. The CNA introduced by Honeycutt and Andersen[53] is one of the frequently used topological methods to interpret the 3D local environment around a reference particle. The algorithm performs geometrical analysis of the nearest neighbors in relation to each reference atom, with a predefined cut off distance for the nearest neighbor shell. However, for multi – phase systems, it is no longer possible to define the global cut – off radius. Therefore, Stukowski *et al.*[51,52] introduced the ACNA method, where the algorithm automatically determines the optimal value of the cut – off radius for each individual particle.

Figure 6(a,b) show the phase fraction variation of FCC and HCP structures as a function of frequency, which correlates well with the bond angle distributions shown in Fig. 5(a,b). For a shaking amplitude of 0.2 mm, both FCC and HCP phase fractions exhibit a maximum at 180 Hz, which is in agreement with the bond angle distribution. In contrast, for a shaking amplitude of 0.6 mm (Fig. 6(b)), the FCC phase fraction shows a maximum at 85 Hz, while the HCP phase fraction increases up to 180 Hz. This behavior is in agreement with the corresponding ADF (Fig. 5(b)), where the evolution of peaks at 109° and 147° indicates the favoring of the HCP phase at higher values of the frequency. A 2D projection of the structural evolution of these different phases (obtained using the ACNA mentioned above) is shown in Fig. 7 for a shaking amplitude of 0.6 mm. As the frequency increases, there is an initial rise in the number of FCC clusters (green spheres in Fig. 7). At 85 Hz, the sphere clusters contain 24.7% atoms in FCC configurations and 13.8% in HCP configurations (red spheres in Fig. 7). A further increase in the frequency, however, causes this trend to be reversed, such that at 180 Hz, the FCC clusters are reduced to 18.4% of the granular assembly, while the HCP phase fraction grows to 22.2%. Videos showing these structural evolutions are attached as supplementary Information.

Finally, Voronoi tessellation analysis has been used to characterize these local structure configurations as a function of frequency (and implicitly packing fraction). The evolution of local structures using Voronoi analysis is often identified using basic Voronoi indices, $<n_3, n_4, n_5, n_6>$, where $n_i$ specify the number of Voronoi faces with i edges in a Voronoi polyhedron. In Fig. 8(a,b) histograms showing the development of most abundant Voronoi polyhedra as a function of frequency for shaking amplitudes of 0.2 and 0.6 mm are depicted, respectively. The Voronoi indices have been categorized as[57,58]:

1. Icosahedral (five – fold) symmetry with the Voronoi indices as $<0, 2, 8, n_6>$, $<0, 1, 10, n_6>$; these are suggestive of random packings.
2. Rectangular and hexagonal (with four- and six-fold symmetry respectively) with the Voronoi indices $<0, 4, 4, n_6>$, $<0, 5, 2, n_6>$; these indicate crystalline clusters
3. Mixed symmetry, indicated by the Voronoi indices $<0, 3, 6, n_6>$.

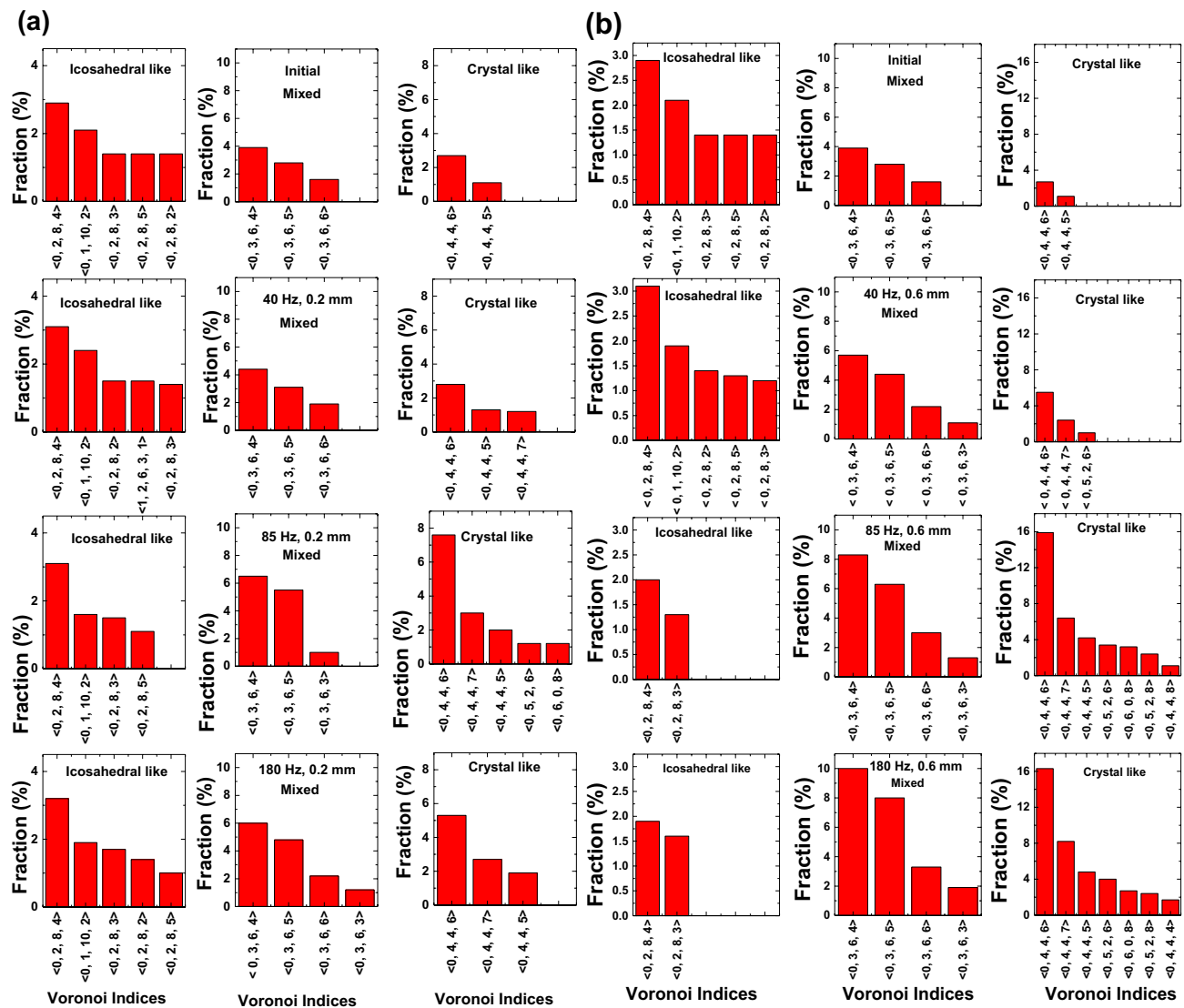

**Figure 8.** Histograms of the Voronoi indices (different polyhedra configurations) as a function of frequency for (**a**) 0.2 mm and (**b**) 0.6 mm shaking amplitudes. Only polyhedra with a percentage greater than 1% are shown.

At the lower shaking amplitude of 0.2 mm in Fig. 8(a), no significant variation in the fraction of Voronoi polyhedra corresponding to icosahedral clusters is observed as a function of increasing frequency. On the other hand and interestingly, the percentage and number of Voronoi polyhedra for crystalline clusters shows an increase at lower frequencies up to 85 Hz, and then reduces thereafter. This reaffirms predictions[33] that at lower amplitudes the crystallization process is incomplete, and it appears that 85 Hz could be the optimum frequency for highest crystalline fraction at this amplitude. At the higher shaking amplitude of 0.6 mm, the fraction of icosahedral clusters remains unchanged and with respect to the data (Fig. 8b) exhibits values similar to that of the lower amplitude data (Fig. 8a). Here however, the fraction of crystalline clusters shows a significant increase, which is also in agreement with the increased packing fraction of 0.71. What appears to be an extremely interesting observation is that while the icosahedral structures appear localized, with their fraction remaining relatively constant even with varying amplitude and frequency, this is by far not the case with the crystalline structures, especially at the higher amplitude, whose fractions increase with increasing frequency. We speculate therefore that the randomly ordered regions may play a rather minor role in the spontaneous transition to crystallization.

Various important conclusions can be drawn on the basis of the above results. First, we highlight the relevance of direct real – space neutron imaging to detect the subtle structural changes resulting from the spontaneous evolution of crystallization from shaken, initially randomly close packed granular assemblies. The present study is the first attempt at using real – space neutron tomography, apart from initial experiments carried out earlier[59], to understand the development of crystalline ordering with detailed local structural analysis, which results from the systematic variation of shaking amplitudes and frequencies with a concomitant rise in packing fraction $\phi$ from 0.62 to 0.71.

We have also demonstrated that granular packings contain localized icosahedral clusters in the disordered initial state of $\phi \approx 0.62$; then, as the shaking amplitude and frequency are increased in combination with increasing packing fraction to $\phi \approx 0.71$, there is a sharp increase in the number of crystalline clusters. In our data, we notice only a small decrease in the fraction of icosahedral clusters as the transition to crystallinity is approached; also there seems to be little correlation between the decrease in the icosahedral clusters and the growth of crystalline clusters. At the lower shaking amplitude of 0.2 mm, we observe that there is no significant variation in the fraction of icosahedral clusters, while the crystalline clusters grow and exhibit a maximum as a function of frequency. This suggests that at lower shaking amplitudes, crystallization is incomplete and/or unstable[33]. The ACNA analysis reinforces these conclusions, with its demonstration of the growth of both FCC and HCP clusters, which also shows a maximum as a function of frequency. In contrast, at the higher shaking amplitude of 0.6 mm, a significant increase in crystalline clusters is observed with increasing frequency. The ACNA analysis shows the presence of both FCC and HCP clusters. There is a competition between these two kinds of ordering[33] which leads to initially, the domination of FCC giving way to the domination of HCP. A possible reason for the dominance of HCP at higher frequencies is that they have inherently higher rigidity. However, it is important to note that the theoretical simulations carried out by Anikeenko et al.[60]. show that as a function of packing fraction large spread in phase fraction of FCC and HCP clusters is observed. This implies that for the same frequency and amplitude of shaking it is possible that different phase fraction values of FCC and HCP clusters are observed, starting from different initial disordered packing state.

The evolution of the crystalline phase as a function of packing fractions in granular sphere packings has also been studied experimentally earlier; in particular, using refractive index matching techniques, Panaitescu et al.[61] and Rietz et al.[62] showed the emergence of clusters with FCC and HCP ordering with cyclically sheared granular spheres. Our study further indicates that at high enough frequencies the HCP phase begins to dominate at the expense of FCC clusters, as shown in Fig. 7(d). The HCP phase initially grows from well defined stacking faults, shown in Fig. 7(c). This is to be compared with the simulations of[33] where, at constant (low) frequencies, the variation of amplitudes leads to a densification of the granular assembly, involving the dominance of FCC clusters overall. We suggest therefore that the dominant kind of crystallinity observed in such spontaneous transitions under shaking depends on a delicate balance of amplitude and frequency in a way that is as yet not fully understood.

The role of icosahedral ordering and the coexistence of two kinds of crystalline ordering (FCC and HCP) are issues that have also been investigated in colloidal systems[14–18,63]. Pusey et al.[14]. showed a clear tendency towards the FCC structural ordering in slowly grown hard colloidal spheres. Leocmach et al.[18] also observed crystalline ordering to be more frequently observed than icosahedral ordering in colloidal systems, with a growth in the former and no noticeable growth in the latter as the glass transition temperature was approached.

This competitive interaction between the coexisting FCC and HCP has also been predicted theoretically by simulation studies[33,64–68]. The Monte Carlo simulation studies reported by Shinde et al.[33] correlate the competitive interaction between FCC and HCP structures with the presence of dislocations between differently ordered domains. The numerical study reported by Klumov et al.[67,68]. shows the competitive fluctuations between FCC and HCP clusters in highly crystalline packings, with the existence of a sharp disorder – order phase transition at $\phi_C \approx 0.64$. Also, icosahedral ordering is observed within the narrow packing fraction range, in agreement with our study. Various other calculations[64–66] reported in the literature have indicated that the very small free energy difference between FCC and HCP structures might explain the coexistence of and dynamical competition between these two crystalline structures.

## Conclusion

In this paper, we present the results of real – space neutron tomography measurements on packings of highly monodisperse granular steel spheres, carried out to study the structural evolution of these packings from an initially disordered state to one containing a high degree of crystallization. For the particular values used in this work, it appears that increasing shaking frequency and amplitude leads to the compaction of the assembly, with a rise in packing fraction $\phi$ from 0.62 to 0.71. The disordered initial state ($\phi \approx 0.62$) is characterized by the presence of localized icosahedral ordering, while at higher packing fractions ($\phi \approx 0.71$), there is a spontaneous evolution of crystallinity[33]. A detailed structural analysis of the latter showed the presence of coexisting FCC and HCP clusters characterized by two interesting features. The first of these concerns the fact that while the icosahedral ordering is localized, with a low population which does not grow, there is a significant increase in crystalline ordering as the transition is approached. The second of these is that initially dominant FCC ordering gives way to HCP ordering as the frequency increases, indicating that there is a dynamical competition at play between these two structures. The nature of this dynamical competition involves a subtle interplay of both amplitude and frequency, as can be seen from a comparison of our results with those of numerical simulations[32,33].

## Acknowledgements

This research used resources at the High Flux Isotope Reactor (HFIR), a DOE Office of Science user facility under contract DE-AC05-00OR22725 operated by the Oak Ridge National Laboratory. This material is based upon work supported by the Office of Basic Energy Sciences, U.S. Department of Energy. The team would like to thank Lou Santodonato and the technical support of HFIR for their help in setting- up the shaker. We also acknowledge Nikolay Kardjilov for his contribution in the initial experimental work carried out on beamline CONRAD in Helmholtz Zentrum Berlin., I.D. would like to thank Daniel Olds and Alexander Stukowski for fruitful discussions. A.M. would like to thank the Max Planck Institute for Mathematics in the Sciences, Leipzig, Germany for their support.

## Author Contributions

I.D., T.C. and A. M. designed the experiment, A. M. provided the theoretical background for the present study, I.D. analyzed the data and wrote the manuscript; S.A.J.K. and T.C. helped with the data analysis and discussed the results; all the authors contributed in reviewing the manuscript and provided fruitful comments towards the improvement of the manuscript.

## Additional Information

**Supplementary information** accompanies this paper at https://doi.org/10.1038/s41598-018-36331-1.

**Competing Interests:** The authors declare no competing interests.

**Publisher's note:** Springer Nature remains neutral with regard to jurisdictional claims in published maps and institutional affiliations.